\input psfig.tex
\documentclass{aa}
\begin{document}

\thesaurus{03 (11.03.1; 11.11.1) }

% According to A&A 304, A5-A17, 1995!

\title{The ESO Nearby Abell Cluster Survey
       \thanks{Based on observations collected at the European Southern
               Observatory (La Silla, Chile)}
      }

\subtitle{VI. Spatial Distribution and Kinematics of Early-- and 
	      Late--type Galaxies}
\author{P.A.M. de Theije and P.~Katgert }
\institute{Sterrewacht Leiden, The Netherlands }

\offprints{P.~Katgert}
\date{Received date; accepted date}

\maketitle
\markboth{P.A.M. de Theije and P. Katgert: The ESO Nearby Abell Cluster 
Survey. VI}{Distribution and kinematics of early-- and late--type
galaxies}

\begin{abstract}

Analysis of the data obtained in the ESO Nearby Abell Cluster Survey
(ENACS) has shown that the space distribution and kinematics of
galaxies {\em with} detectable emission lines in their spectra differ
significantly from those of galaxies {\em without} emission
lines. This result, and details of the kinematics, were considered as
support for the idea that at least the spirals with emission lines are
on orbits that are not isotropic. This might indicate that this subset
of late--type galaxies either has `first approach'-orbits towards the
dense core of their respective clusters, or has orbits that `avoid'
the core.

The galaxies with emission lines are essentially all late--type
galaxies. On the other hand, the emission-line galaxies represent only
about a third of the late--type galaxies, the majority of which do not
show detectable emission lines. The galaxies without emission lines
are therefore a mix of early-- and late--type galaxies. In this paper
we attempt to separate early-- and late--type galaxies, and we study
possible differences in distribution and kinematics of the two galaxy
classes.

For only about 10\% of the galaxies in the ENACS, the morphology is
known from imaging. Here, we describe our classification on the basis
of the ENACS spectrum. The significant information in each spectrum is
compressed into 15 Principal Components, which are used as input for
an Artificial Neural Network. The latter is `trained' with 150 of the
270 galaxies for which a morphological type is available from
Dressler, and subsequently used to classify each galaxy. This yields a
classification for two-thirds of the ENACS galaxies.

The Artificial Neural Network has two output classes: early--type
(E+S0) and late--type (S+I) galaxies. We do not distinguish E and S0
galaxies, because these cannot be separated very robustly on the basis
of the spectrum.  The success rate of the classification is estimated
from the sample of 120 galaxies with Dressler morphologies which were
not used to train the ANN. The success rate is higher for early--type
than for late--type galaxies ($78\pm6\%$ vs. ${63\pm6\%}$). The
weighted average success rate, irrespective of type, is
${73\pm4\%}$. The success rate is somewhat larger for the training
set, and highest for the galaxies with emission lines.

Of the 3798 galaxies that were classified from their spectrum
${57\pm7\%}$ are of early type, and ${43\pm7\%}$ of late type. Using a
subset of these 3798 galaxies, we constructed a composite cluster of
2594 galaxies, 399 of which have emission lines and are therefore
almost exclusively spirals and irregulars. The kinematics and spatial
distribution of the late--type galaxies {\em without emission lines}
resemble much more those of the early--type galaxies than those of the
late--type galaxies {\em with emission lines}. Yet, the late--type
galaxies without emission lines may have a somewhat larger velocity
dispersion and a slightly less centrally concentrated distribution
than the early--type galaxies.

Only the late--type galaxies {\em with emission lines} appear to have
a considerably larger global velocity dispersion and a much less
concentrated projected density profile than the other galaxies. Thus,
the suggestion of fairly radial, and possibly `first approach' orbits
applies only to spirals with emission lines. The early--type galaxies
with emission lines (among which the AGN), may also have a large
velocity dispersion and be concentrated towards the cluster centre.

\end{abstract}


\section{Introduction}
\label{introduction}

Recently, the kinematics of the different types of galaxies in
clusters, as well as their spatial distribution, has received some new
attention, e.g. from Colless \& Dunn (1996), Carlberg et al. (1996)
\& Mohr et al. (1996). Generally speaking, the late--type galaxies
are found to avoid the central regions of their clusters, while their
line--of--sight velocity dispersion with respect to the average
velocity of the cluster, ${\sigma_{los}}$, appears to be higher than
that of the early--type galaxies.

The ESO Nearby Abell Cluster Survey (ENACS, Katgert et al. 1996, 1998)
has yielded redshifts for more than 5600 galaxies in the directions of
about 100 rich clusters of galaxies, mostly in a cone around the South
Galactic Pole. Using the ENACS, Biviano et al. (1997, hereafter Paper
III) compared the spatial distribution and kinematics of the galaxies
with emission lines (hereafter: ELG) with those of the galaxies
without emission lines (hereafter: non--ELG). From a subsample of 545
galaxies for which a morphological type is known, Biviano et
al. concluded that the ELG are almost exclusively late--type galaxies,
i.e., spirals (and irregulars), while the non--ELG are a mix of
early-- and late--type galaxies. They found that ${\sigma_{los}}$ of
the ELG is, on average, 20\% larger than ${\sigma_{los}}$ of the
non--ELG, while the spatial distribution of the ELG is significantly
less peaked towards the cluster centre than it is for the
non-ELG. These two facts, in combination with details of the
kinematics, were interpreted as evidence for a picture in which the
ELG are mostly on fairly radial, 'first approach' orbits towards the
central regions of their clusters, and thus not in full equilibrium
with the population of non--ELG.

Biviano et al. estimated that the large majority of the ELG are
spirals; however, the ELG appear to represent only about one--third of
the total spiral population. Therefore, it was not clear whether the
conclusion about the difference in kinematics of ELG's applies only to
spirals with emission lines, or whether it applies to all spirals. If
the spirals among the non--ELG would have identical kinematics and
spatial distribution as the spirals with emission lines, the real
differences between early-- and late--type galaxies would be larger
than the apparent differences. That would bring the result of Biviano
et al. more in line with that of Colless \& Dunn (1996) who found
that the velocity dispersion of the late--type galaxies in the main
concentration of the Coma cluster is very close to ${\sqrt{2}}$ times
that of the early--type galaxies.

On the other hand, it is conceivable that only the spirals with
emission lines are on 'first approach' orbits, which would be
consistent with the presence of sufficient amounts of line--emitting
gas. The spirals without emission lines might then have traversed the
central regions of their clusters and have lost most of their
line--emitting gas in the process.

Recently, Ram\'{\i}rez \& de Souza (1998) concluded that the orbits
of elliptical galaxies in clusters are close to radial, while spirals
have more circular shaped, or isotropic, orbits. Their conclusion is
based on an analysis of the distribution of line-of-sight velocities.

In order to be able to investigate the kinematics of the spirals
without emission lines, as well as to elucidate the cause for the
apparent disagreement between the results of Ram\'{\i}rez \& de Souza
and that of Biviano et al., we need morphological types for the
(non--ELG) galaxies in the ENACS. The obvious way to get these is
through imaging. With well over 4000 galaxies without emission lines,
that represents a major observational effort, on which we have
embarked, but which will take some time to finish.

In this paper, we adopt a different approach, by using the ENACS
spectra. The morphological types are estimated from the spectra with a
Principal Component Analysis, in combination with an Artificial Neural
Network. The network is 'trained' with a subset of the ENACS galaxies
for which a morphological type is available from imaging (Dressler,
1980, hereafter D80) and it is `tested' with the remaining galaxies
with morphology from Dressler. With the morphological types estimated
from the spectra, we investigate the kinematics and spatial
distribution of, in particular, the late--type galaxies with and
without emission lines.

The paper is organized as follows: in Sect.~\ref{data} the ENACS data is
summarized. In Sect.~\ref{methods} we describe the algorithm that we used
to estimate the morphological type of a galaxy from its spectrum, by
applying a PCA and an ANN. In Sect.~\ref{results} we discuss the results
of the combined PCA/ANN and present the success rates achieved in
assigning morphological types. In Sect.~\ref{spatkin} we present an
analysis of the spatial and kinematical differences between the
(subets of) early-- and late--type galaxies. In Sect.~\ref{conclusions}
we summarize the results.

\section{Data}
\label{data}

In the present analysis we use the galaxy spectra obtained with the
OPTOPUS instrument at the ESO 3.6m telescope at La Silla, Chile in the
context of the ENACS. For a detailed description of the
characteristics of the survey, we refer to Katgert et al. (1996,
1998). A brief summary of those aspects of the observations that are
relevant to the analysis in this chapter may be useful, however.

The observations were done between September 1989 and October 1993.
The observed galaxies all lie in the direction of rich Abell clusters.
The redshifts of these clusters are mostly ${\le 0.1}$ and most
clusters lie around the South Galactic Pole in the solid angle defined
by ${b \le -30^{\circ}}$ and ${-70^{\circ} \le \delta \le 0^{\circ}}$.
The galaxies were selected either on film copies of the SERC IIIa-J
survey or on glass copies of the first Palomar Sky Survey. Areas of
between 1 and 4 square degrees centered on the target clusters were
scanned with the Leiden Astroscan plate--measuring machine. The
magnitude limits are between 16.5 and 17.5 in the R--band (see Katgert
et al. 1998 for details). These limits correspond to absolute
magnitudes of --19.8 and --18.8 at the median redshift ${z = 0.06}$ of
the survey, assuming a Hubble parameter of 100 km s${^{-1}}$
Mpc${^{-1}}$.

The OPTOPUS system used fibres with a diameter of 2.3 arcseconds,
which corresponds to a linear scale of 2.1 $h^{-1}$ kpc at a redshift
of ${z = 0.06}$. In 6 of the 9 observing runs the same spectrograph
setup was used. In general, the wavelength range was from ${\approx
3850}$ \AA \ to ${\approx 6000}$ \AA, but it varies slightly between
runs. The spectral resolution is almost always 130 \AA/mm, or about 5
\AA, except in the run of September 1989, which has a lower
resolution. Due to different pixel sizes of the CCD detectors used,
the spectra were sampled at either 1.9 or 3.5 \AA/pixel.

In the wavelength range covered by the observations, and for the
redshifts of the clusters observed, the principal emission lines that
were observable are [OII] (3727 \AA), H${\beta}$ (4860 \AA) and the
[OIII] doublet (4959 and 5007 \AA). Possible emission lines were
identified independently by two persons: in the 2--D frames and in the
1--D extracted spectra (see Paper III for details). In the spectra of
about 1200 galaxies one or more emission lines were detected. For
554\footnote{This number is slightly larger than the number mentioned
in Paper III (which was 545), due to an updated cross--reference
between the ENACS sample and the Dressler catalogue.} galaxies in the
10 ENACS clusters in common with the sample of Dressler (1980) the
morphology is available. Of the 71 ELG which have a morphological
classification in D80, 61 are spirals or irregulars (86\%), 8 are S0's
(11\%) and 2 are ellipticals (3\%). On the other hand, of the 181
spirals that the D80 set has in common with ENACS, 61 show emission
lines. So, while 6 out of 7 ELG are spirals/irregulars, only 1/3 of
the spirals are ELG. A small fraction (about 7\% according to Biviano
et al. 1997) of the ELG are active galactic nuclei, as is evident from
the line widths.

The spectra that were obtained in September 1992 often exhibit
peculiarities, such as deviant pixels at beginning and end of the
spectrum. These probably were introduced in the reduction process and
have not influenced the redshift determination.  We have not reduced
these spectra again for the present analysis, but we have excluded
them from the analysis, as they could produce below-average
classification results (see Sect.~\ref{sucrat}).

\section{Spectral Classification}
\label{methods}

For the classification of the galaxies on the basis of their spectrum,
we use a two--step scheme in which we first describe a spectrum in
terms of its most significant Principal Components (PCs), and then
use a trained Artificial Neural Network to classify the galaxy on the
basis of those components. In this section, we summarize the methods
that we used and the details of their implementation as far as those
are required for an appreciation of the results.

Several authors applied PCA (either by itself or in combination with
an ANN) to the problem of trying to determine from a spectrum the
stellar or galaxy type. Deeming (1963), in classifying stellar
spectra, found a very good correlation between the first, most
important PC and the stellar type. Francis et al. (1992) carried out a
study of a large sample of QSO spectra, and developed a classification
scheme based on the first three PCs.  Von Hippel et al. (1994) used an
ANN to classify stellar spectra and concluded that they could recover
the stellar type to within 1.7 spectral subtypes. Sodr\'e Jr. \&
Cuevas (1994) showed that the spectroscopic parameters extracted from
the spectra of galaxies, like the amplitude of the 4000 \AA \ break or
of the CN band, correlate well with Hubble type.

Zaritsky et al. (1995) decomposed galaxy spectra into an old stellar
component, a young stellar component and various emission-line
spectra. They classified the galaxies by comparing the relative
weights of the components with those of galaxies of known
morphological type and found that the spectral classification agreed
with the morphological classification to within one type (e.g. E to S0
or Sa to Sb) for ${\ge 80\%}$ of the galaxies. Connolly et al. (1995)
decomposed each spectrum into eigenspectra and found that the
distribution of spectral types can be well described by the first two
eigenspectrum coefficients.

Folkes et al. (1996) combined PCA and ANN to classify galaxy spectra.
Their purpose was to investigate galaxy classification from spectra to
be obtained in the 2dF Galaxy Redshift Survey. They generated
artificial spectra and obtained a success rate of more than 90\% in
recovering the galaxy type from the spectrum.

Lahav et al. (1996) used ESO--LV galaxies (Lauberts \& Valentijn 1989)
and grouped them in three ways. From a PCA applied to 13 galaxy
parameters they found that different morphological types occupy
distinct regions in the plane defined by the two most important
PCs. They also used an ANN with the 13 galaxy parameters as input and
concluded that with a single output node, there is a strong
correlation between the galaxy type indicated by the ANN--output and
the input type. Using two output nodes, one for early-- and one for
late--type galaxies, the overall success rate was 90\%, which
decreased to 64\% if 5 output nodes were used, viz. E, S0, Sa+Sb,
Sc+Sd and I.

In the last few years several more applications of PCA analysis,
combined PCA and ANN analysis, or ANN analysis were published. In the
field of stellar classification see e.g. Bailer-Jones (1997), Ibata
\& Irwin (1997), Weaver \& Torres-Dodgen (1997) and Singh et
al. (1998), and in the area of galaxy classification e.g. Galaz \& de
Lapparant (1998) and Bromley et al. (1998).

\subsection{Principal Component Analysis}
\label{pca}

Principal Component Analysis (PCA) is a technique developed for data
compression as well as data analysis. As measured parameters, like e.g
spectral fluxes, may be correlated, it is of interest to determine the
minimum number of {\em independent} variables that can describe the
larger amount of correlated observed parameters. A full 
description of PCA can be found in e.g. Kendall \& Stuart (1968) and
Folkes et al. (1996). In our analysis, the PCA is an important first
step as it reduces the number of parameters that describe the galaxy
spectrum, while it recovers essentially all significant information
and reduces the noise.

\subsubsection{Preparing the data}
\label{prep}

Before we could apply PCA to the ENACS galaxy spectra some
preparations were necessary. First, all spectra were inspected and a
few spectra with strong discontinuities or other non--physical
features were discarded. Secondly, sky lines were removed by linear
interpolation. Thirdly, the spectra were shifted back to
zero--redshift and corrected for the response functions of the OPTOPUS
instrument (spectrograph and CCD detector). Fourthly, a maximum common
(zero--redshift) wavelength range had to be established for as large a
subset of the galaxy sample as possible.

Using all galaxies in the ENACS survey, this common wavelength range
would be rather small because background galaxies have redshifts up to
${z \approx 0.15}$. We have chosen to use the zero--redshift
wavelength range from ${\lambda_{min} = 3720}$ \AA \ to
${\lambda_{max} =}$ 5014 \AA.  This range includes all 4 major
emission lines (see Sect.~\ref{data}) and provides at least 7 \AA \
continuum beyond the [OII] 3727 \AA \ and [OIII] 5007 \AA \ lines. All
spectra were resampled in the range [${\lambda_{min}, \lambda_{max}}$]
with ${\Delta \lambda / \mbox{pixel} = 3.5}$ \AA, which yields 371
spectral fluxes.

For some field galaxies the zero-redshift spectrum did not fully cover
the wavelength range [${\lambda_{min},\lambda_{max}}$]. When the
wavelength coverage of a galaxy spectrum fell short by more than 70
\AA \ (i.e., 20 pixels) from the [${\lambda_{min}, \lambda_{max}}$] 
interval, the galaxy was removed from the sample. When the galaxy
spectrum fell short by less than 70 \AA, it was extrapolated by a
second--order polynomial either down to ${\lambda_{min}}$ or up to
${\lambda_{max}}$, or both. This extrapolation does not introduce
major errors in the fluxes at the edges of the spectrum.

Finally, the spectra were normalized to unit integral. For the
normalization we interpolated the spectrum in the regions of 20 \AA \
centered on the emission lines because a strong emission line may
result in a continuum which is too low.

Leaving out all spectra that were observed in September 1992 and
rejecting all galaxies for which more than 20 pixels had to be added
at either one or both ends to fill the spectral range
[${\lambda_{min},\lambda_{max}}$], we retained 3798 galaxies for the
PCA. For 270 of these, a morphological classification is available
from D80.

\begin{figure}
\psfig{figure=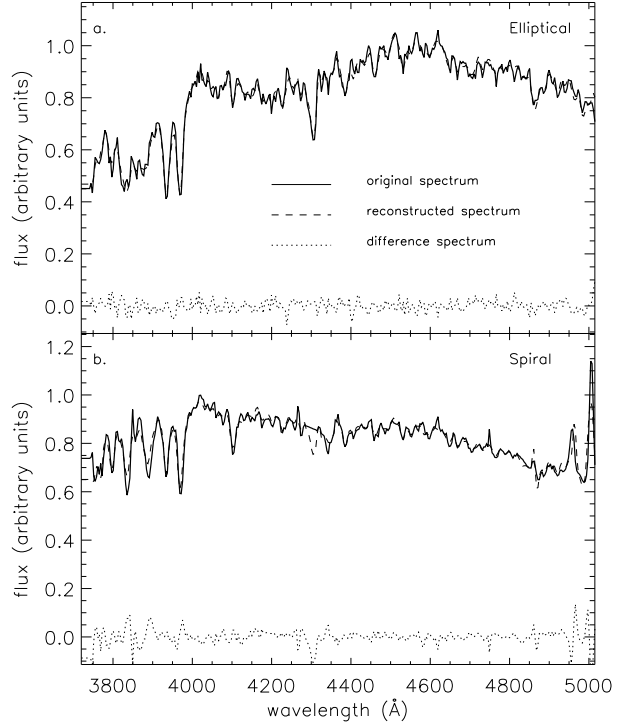,width=8.6cm}
\caption{\label{reconstructed}
Comparison of two ENACS spectra with the reconstruction of the same
spectra from the first 15 Principal Components. The solid line is the
observed galaxy spectrum, the dashed line is the reconstructed
spectrum. The dotted line indicates the difference between the
observed and the reconstructed spectrum.  {\bf a:} Elliptical galaxy,
{\bf b:} Spiral galaxy.}
\end{figure}

\subsubsection{Determining the Principal Components}
\label{appl_pca}

After the preparation described in Sect.~\ref{prep} the resampled
spectrum of each galaxy defines a j--dimensional vector ${\vec{x}}$,
whose components represent the flux in the j pixels of the spectrum,
with j = 1--371. From each component ${x_{j}}$ we subtract the mean
over all galaxies, ${\overline{x_{j}}}$, to centre the j--th parameter
on zero (remember that we normalized all spectra to the same integral
of 1.0). The values ${x_{j}-\overline{x_{j}}}$ can be used in two
different ways in the PCA. Firstly, ${x_{j}-\overline{x_{j}}}$ may be
normalized by its standard deviation ${\sigma_{j}}$. In that case,
each of the components of the spectral vectors has unit variance for
the set of spectra used. This method is sometimes recommended as it
puts each input parameter on a similar scale. In this way, one may
construct vectors from components that are not related, such as
e.g. mass and size.

However, there are cases in which the different dispersions or the
relative strengths of the inputs are important (see e.g. Folkes et
al. 1996). E.g., in our PCA the components of the spectral vectors
that contain the principal emission lines will have a larger variance,
as these may or may not be present, and it may be important to retain
this information. We did the PCA with and without normalization with
the standard deviation, and obtained better results with normalization
than without.

The PCA solves for the weights $w_{kj}$ that define the 371 PCs $e_k$
which follow from the spectral fluxes by the relation: ${e_{k} =
\sum_{j=1}^{371} w_{kj}(x_{j}-\overline{x_{j}})/\sigma_{j}}$. The PCs 
are thus linear combinations of the normalized spectral fluxes and
form an orthogonal basis. The first PC, ${e_{1}}$, contains most of
the variance between the spectra and describes the most characteristic
difference between the spectra. The last PC contains least of the
variance and will be most affected by noise. In practice, we have
restricted the ANN analysis to the 15 most significant PCs (see 
Sect.~\ref{sucrat}).

In Fig.~\ref{reconstructed} we show two examples of spectra and
their PCA reconstruction, based on the first 15 PCs. These examples
illustrate that the spectra can be reconstructed quite well with only
a limited number of PCs. There may be some indication that the
spectrum of an elliptical is easier to represent with 15 PCs than that
of a spiral, but the difference is slight.

\subsubsection{Physical meaning of the PCs}
\label{meaning}

\begin{figure}
\psfig{figure=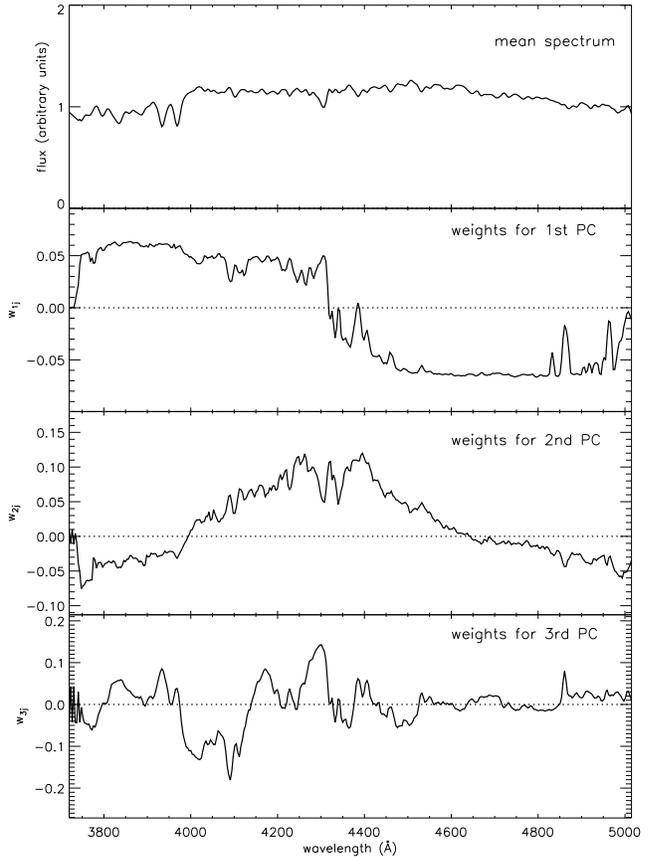,width=8.6cm}
\caption{\label{pcs}
Average galaxy spectrum, and the weights $w_{ij}$ for the first 3
Principal Components (i=1--3), calculated for the entire data set of
3798 galaxies. See text for more details.}
\end{figure}

In Fig.~\ref{pcs} we show the average spectrum of the 3798 galaxies in
the data set, together with the weights for the 3 most significant
PCs, i.e., ${w_{1j}}$, ${w_{2j}}$ and ${w_{3j}}$. The weights for the
first PC, ${w_{1j}}$, indicate that ${e_{1}}$ represents the colour of
the galaxy, as it measures the flux in the interval ${\lambda =
[3720,4350]}$ \AA \ minus the flux in the interval ${\lambda =
[4350,5120]}$ \AA. The wavelength dependence of ${w_{1j}}$ is very
similar to that in Sodr\'e \& Cuevas (1997, their figure 5). The
second PC, with weights ${w_{2j}}$, apparently measures the curvature
of the spectrum, i.e. the flux between ${\lambda = 4000}$ \AA \ and
${\lambda = 4600}$ \AA, minus the flux below and above these
wavelengths. The weights for the third PC, ${w_{3j}}$, have a
signature just redwards of the 4000 \AA \ break, and $e_3$ thus seems
to be sensitive to the strength of this break. Sodr\'e \& Cuevas
(1994) noted that the 4000 \AA \ break correlates well with
Hubble--type. The third PC also appears to weigh the G--band at
${\lambda \approx 4300}$ \AA. It gets progressively more difficult to
understand in detail the physical meaning of the higher order PCs, but
they gauge the various less conspicuous features in the spectrum, such
as the many absorption and emission lines.

Note that we expressed the PCs in terms of the spectral data, which
provides an immediate physical meaning of the weights ${w_{ij}}$. In
an alternative but fully equivalent representation, the spectral data
can be approximated by a weighted sum of eigenspectra; see e.g.
Connolly et al. (1995) or Galaz \& de Lapparent (1998).

\subsection{Artificial Neural Networks}
\label{ANN}

The first 15 PCs derived for each spectrum are used as input for an
Artificial Neural Network (ANN).The ANN determines the optimum way of
combining the PCs in order to obtain a single number which maps, with
maximum discriminating power, onto the desired quantity which, in our
case, is morphological type. ANN's are frequently used to recognize
patterns in input data. An array of parameters is presented as input
to the ANN, which must have been trained to recognize the desired
patterns. The ANN then yields the class of object for which the input
array is most characteristic. The classification is objective: the ANN
is true to the training it received, and repeatable.

An ANN uses weights to translate the input data into one or more
parameters which can be compared with the corresponding parameters for
the training set in order to estimate the class of an object. The
weights in the ANN are determined by an iterative least--squares
minimization using a back--propagation algorithm. In each iteration
step, the current values of the weights are updated according to the
difference between the supplied output type and the calculated output
type. For a full description of ANN's, the reader is referred to
e.g. Hertz et al.  (1991), Kr\"ose \& van der Smagt (1993) and Folkes
et al. (1996).

\subsubsection{Training the ANN and tuning its parameters}
\label{training}

We trained the ANN by using the spectra of 150 of the 270 galaxies in
our sample of 3798 for which Dressler (1980) gives a morphological
type. The median redshift of the clusters studied by Dressler is about
0.04, which is significantly smaller than the median redshift of the
ENACS sample of about 0.07. The 10 clusters in common between D80 and
ENACS have redshifts between 0.04 and 0.07. The training set contains
approximately equal numbers of galaxies in each of the three
morphological classes that we attempted to `resolve', viz. E, S0 and
S+I.

The complexity of an ANN depends strongly on the number of inputs per
galaxy and on the number of hidden nodes, layers, outputs, and
connections. Therefore, one is well advised to use as few of these as
possible (de Villiers \& Barnard 1993), as long as the discriminating
power of the ANN is not affected. For that reason, only the 15 most
significant PCs of the galaxy spectra were presented to the ANN,
rather than all 371 original spectral fluxes. By using only the 15
most significant PCs, we also reduce the noise considerably, as the
latter is mostly contained in the higher--order PCs.

We used only one hidden layer, which contains 5 nodes. This makes the
backpropagation network much more rapid to train (see e.g. de Villiers
\& Barnard 1993).

\begin{figure}
\psfig{figure=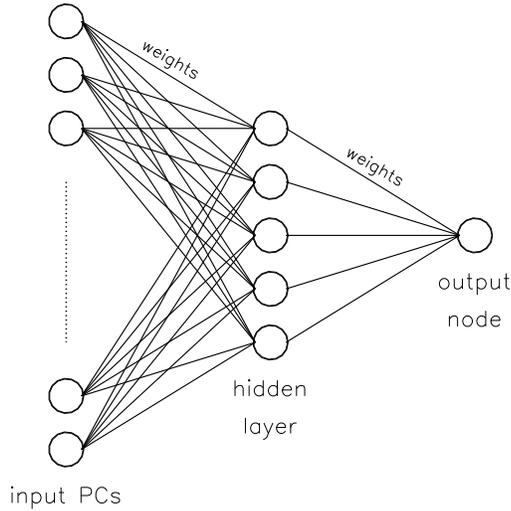,width=8.0cm}
\caption{\label{ANNdiagram}
Schematic diagram of the Artificial Neural Network that we used. The
network determines the galaxy type ('output node'), from the Principal
Components describing the galaxy spectrum. Each node in a given layer
is connected to all nodes in the adjacent layers by weight vectors.}
\end{figure}

Only one output node was used, with output values in the interval
[0,1]. Some authors define a separate output node for each of the
morphological types that can be assigned to the galaxies (e.g.
Storrie--Lombardi et al. 1992). The output node which has the highest
'activity' then determines the galaxy morphology. However, because
galaxies are thought to form a continuous instead of a discrete
sequence of different morphologies (e.g. Naim et al. 1995), we have
chosen to describe the sequence with only one output node with a
continuous range of output values. A schematic diagram of the ANN we
used is shown in Fig.~\ref{ANNdiagram}.

When training the ANN, it is essential to stop the iterative
minimization at the right moment. One option is to stop when the total
error between calculated output types and supplied output types of the
{\em training set}, the so--called 'cost function', drops below a
certain value, or changes little between successive iteration steps.
However, this may result in 'over--fitting', i.e., one may interpret
the statistical fluctuations in the training set as global
characteristics. Another option is to minimize the cost function as
calculated for the {\em test set} (Lahav et al. 1996). Because the
ANN is not trained on this set, the cost function will usually have a
true minimum at a certain iteration step, and increase after that. In
our case the results are almost identical for both options.

As we want to extend the analysis in Paper III to early-- and
late--type galaxies, we are primarily interested in a two--class
classification. Therefore, we trained the ANN for a pure
early--/late--type division which allows a separation of the
heterogeneous class of non--ELG into early-- and late--type
galaxies. An additional reason for taking ellipticals and S0's
together in one class was given by Lahav et al. (1996), who found that
76\% of all early--type galaxies were correctly classified by their
ANN, but that of the S0's only 66\% was classified correctly. They
suggested that this may be an indication that the S0's form a
`transition class' in the Hubble sequence. Sodr\'e \& Cuevas (1997)
found that the first, most significant PC of ellipticals and S0's are
very similar, so it is hard to distinguish between them on the basis
of their spectrum.

We have also trained the ANN for a three--class division into E, S0
and S+I. We defined the output values of the ANN for these three
classes to be 1/6, 1/2 and 5/6. In principle, we could have defined
different output values for these three categories which would have
resulted in different weights in the trained ANN. However, we find
that an ANN with output values of 0, 1/2 and 1 gives classification
results that are essentially identical to those obtained with the
output values 1/6, 1/2 and 5/6. Galaxies are assigned the
morphological classification for which the difference between their
ANN output parameter and the output parameter defined for the class is
smallest. After running the three--class ANN we also sum the E and S0
categories to produce the equivalent of the early--type category in
the two--class ANN. We find that there are no significant differences
between the results of a true two--class ANN and a semi two-class ANN
obtained by combining the E and S0 classes in a three--class ANN.
Below we will describe the results for the latter.

\subsubsection{Testing the ANN}

In addition to training the ANN, we tested it with a test set
consisting of the remaining 120 galaxies with morphology from D80. The
results for this test set, in terms of the success in classifying the
galaxies correctly, are valid for the entire data set of ENACS
galaxies for which no morphological classification is available.
However, for the latter we do have information on the presence or
absence of detectable emission lines in the ENACS spectrum, and we
will use that to refine the determination of the ANN output parameter
which best separates early-- and late--type galaxies.

\subsubsection{Optimizing the classification results}
\label{optres}

Our goals are to optimize the success rate of the classification, to
obtain the observed fraction of late--type galaxies among the ELG
(viz. 86\%, see Paper III), and to obtain the correct fractions of E,
S0 and S+I galaxies used to train and test the ANN. The only freedom
one has to achieve these goals, after tuning all parameters as
described in Sect.~\ref{training}, is to set the output ranges within
which a galaxy will be classified as E, S0 or S+I. A priori, the most
logical choice the output ranges is [0,1/3], [1/3,2/3], and [2/3,1]
for E, S0 and S+I, respectively. However, we find that the ranges
[0.00,0.34], [0.34,0.59] and [0.59,1.00] produce a fraction of
early--type galaxies among the ELG that is more consistent with
observations than that found with the a priori choice. In the
following we will therefore use the ranges [0.00,0.34], [0.34,0.59]
and [0.59,1.00]. Note, however, that the success rates for the two
sets of output ranges differ by at most a few percent.

\subsection{Possible causes of misclassification}
\label{misclass}

There are a number of factors that determine the performance of the
classification algorithm.

First, the representation of the spectra by the first 15 PCs is not
perfect. However, the error one makes if one only uses the first 15
PCs is probably quite small (see Fig.~\ref{reconstructed}), while
the results are not expected to depend much on the exact number of PCs
used, as long as this number is large enough (see also
Sect.~\ref{sucrat}). On the other hand, it is likely that the
correspondence between the characteristics of the spectrum (as
quantified in the first PCs) and the morphology is not entirely
one-to-one. For instance, the spectrum of a late--type galaxy is
likely to depend on the location in the galaxy; viz., if the aperture
of the spectrograph covered only the central region of such a galaxy,
a significant contribution of a bulge may well create an apparent
inconsistency between morphological and spectral classification.

\begin{table}
\begin{tabular}{lcccc}
\hline
\multicolumn{5}{c}{Three--class system} \\
\hline
    &    E &   S0 &  S+I &   \% \\
\hline
E   &   15 &    1 &    0 & 0.94 \\
S0  &    3 &   11 &    2 & 0.69 \\
S+I &    1 &    1 &    6 & 0.75 \\
\hline
\%  & 0.79 & 0.85 & 0.75 &      \\
\hline
\hline
    &      &      &      &      \\
\hline
\multicolumn{5}{c}{Two--class system} \\
\hline
     & \multicolumn{2}{c}{E/S0} &  S+I &   \% \\
\hline
E/S0 & \multicolumn{2}{c}{30}   &    2 & 0.94 \\
S+I  & \multicolumn{2}{c}{2}    &    6 & 0.75 \\
\hline
\%   & \multicolumn{2}{c}{0.94} & 0.75 &      \\
\hline
\hline
\end{tabular}
\hfill
\caption{\label{Dr_intern} 
Distribution of morphological type for the galaxies that Dressler
(1980) classified twice, in the cluster DC 0326--53 as well as in DC
0329--52. The last column and the bottom line of each half of the
Table indicate the fraction of galaxies that is classified into the
same class twice.}
\end{table}

Secondly, morphological classification is not easy. For example, it is
likely that some S0 galaxies, especially those seen face--on, are
classified as ellipticals, on the basis of the image only (this is
less likely if the brightness profile is used as well). On the other
hand, edge--on SO galaxies and spirals are not always easy to classify
correctly. Naim et al. (1994) showed that there is indeed some
ambiguity in classifying galaxies solely on the basis of the
morphology of the images. They found 6 experts willing to classify a
set of 831 galaxy images. The results show that both types of
disagreement mentioned above do indeed occur, as well as differences
in the verdict of whether a spiral galaxy is of early or late
type. The {\em r.m.s.} differences between verdicts of experts ranged
from 1.3 to 2.1 Revised Hubble types. This is as large as the {\em
r.m.s.} dispersion between the {\em mean classification} of the 6
experts on the one hand and the results of an ANN analysis on the
other (Naim et al. 1995).

Even an expert is not always totally consistent. Using the 40 galaxies
in Dressler's catalogue that were classified twice, once in the
cluster DC 0326--53 and once in DC 0329--52, this effect can be
quantified. A comparison between both classifications of D80 is given
in Table \ref{Dr_intern}. In the three--class system, 8 out of the 40
galaxies have an inconsistent classification, and at least 4 of the 40
classifications (or 10\%) are thus incorrect. In addition, it cannot
be excluded that galaxies for which both independent classifications
are identical, have yet been classified incorrectly; so, the 10\% of
misclassifications is a lower limit. If one takes E and S0 galaxies
together to obtain an early-- vs. late--type classification, the
number of misclassifications is at least 5\%. Note, however, that the
other way to read this number is that D80 is close to 95\% consistent:
an impressive achievement, as will be confirmed by anybody who has
done morphological classification.

Thirdly, as mentioned before, the spectral difference between E and S0
galaxies probably is not very large (see Lahav et al. (1996) and
Sodr\'e \& Cuevas (1997)).

Fourthly, Zaritsky et al. (1995) found that for 51 of the 304 galaxies
in their sample (i.e. for 17\%) the spectral typing is not consistent
with the morphological classification to within one morphological type
(E, S0, Sa, Sb, Sc and Irr). In 36 cases there is a discrepancy
between morphological and spectral classification that transgresses
the early--/late--type galaxy boundary. It is noteworthy that there
are 16 cases of early--type morphology with late--type spectrum
(mostly on the basis of emission lines), and 20 cases of late--type
morphology with early--type spectrum. I.e. the effects of
misclassification appear to be more or less symmetric, and can
therefore be considered as sources of random errors, like the effects
mentioned above.

For several hundred of the ENACS galaxies that we studied in this
paper and for which we obtained a spectral classification, we also
have obtained CCD images which yield a morphological classification.
Provisional results indicate that for our spectral classification
method, the misclassification probably is not symmetric between
early--type morphology/late--type spectrum confusion and vice versa
(Thomas \& Katgert, in preparation). Only 10\% of the E and E/S0
galaxies in our sample have a $>$50\% probability that their spectrum
is late--type. For the S0 galaxies this fraction increases to about
20\%. However, about 30\% of the spiral galaxies has a spectrum that
has a $>$50\% chance of being indicative of an early--type galaxy.
This suggests that the chance of a spectral misclassification of an
early-type galaxy is considerably smaller than that of a spectral
misclassification of a late--type galaxy. Presumably, the fact that
the ENACS spectra sample only the central few kpcs of the galaxies,
amplifies the influence of the bulges on the spectral classification
of late--type galaxies.

\subsection{Dependence of results on algorithm parameters}
\label{parinf}

A number of choices were made and parameters were chosen in our
classification algorithm. The first one is the number of PCs that is
used in the ANN. In principle, this number is important, as using too
few components to describe the spectrum will make the fits to the
original spectra less accurate. The ANN will then have more problems
to classify the galaxies. If one uses too many components, one may
model the spectra too precisely and use PCs which are too noisy. In
Sect.~\ref{sucrat} we will check how the results depend on the
number of PCs.

\begin{figure*}
\psfig{figure=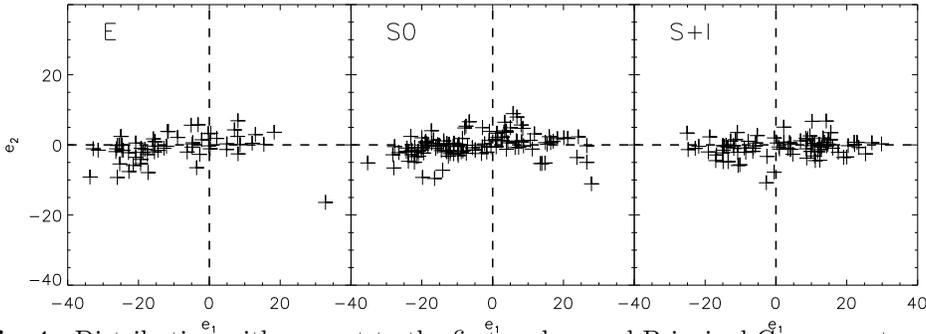,width=12.0cm}
\caption{\label{e1e2_1}
Distribution with respect to the first and second Principal
Components, ${e_{1}}$ and ${e_{2}}$, for galaxies of
type E (62 galaxies), S0 (118) and S+I (90), as classified by D80.}
\end{figure*}

If one does not normalize the spectral fluxes to unit variance (see
Sect.~\ref{appl_pca}), the relative strengths of the spectral points
are retained. This may be important (FLM96) and it emphasizes certain
emission or absorption lines. If we do not normalize to unit variance,
the success rates are smaller than when all pixels are normalized to
unity. Sodr\'e \& Cuevas (1997) also mention that the spread in the
first two PCs is larger if the input parameters are all normalized to
unit variance.

The exact number of nodes in the hidden layer of the ANN is of minor
importance. Using 5 or 7 nodes gives essentially the same results, but
using only 3 nodes produces results that are slightly worse. The
exact values of the learning parameters and the weight--decay term of
the ANN (see e.g. Kr\"ose \& van der Smagt 1993) are not important
either, as long as they are sufficiently small, i.e., 0.001--0.01.
The number of cycles through the input list of training galaxies that
is needed, however, does depend on the values of the learning
parameters.

The number of galaxies used to train the ANN is not critical, as long
as it is sufficiently large, say 150. However, the {\em r.m.s.}
values around the average classification result (see Sect.~\ref{sucrat})
may depend on the number of galaxies in each of the morphological
classes. The number of galaxies with which we have chosen to train the
ANN, viz. 150, is a compromise between having a sufficient number of
galaxies to train the ANN, and having enough galaxies left to test the
performance of the ANN. It is important, however, that all three main
morphological types are represented in the training set with roughly
equal numbers. If one morphology is overrepresented with respect to
the other types in the training set, there will be a positive bias for
that particular type. So, in principle, the composition of the
training set should closely mimic the composition of the sample to be
classified in order to have minimum bias.

\begin{table}[b]
\begin{tabular}{lrcc}
\hline
type &    N & \%(${e_{1}>0}$) & \%(${e_{2}>0}$) \\
\hline
E    &   62 &  24           &  40           \\
S0   &  118 &  36           &  45           \\
S+I  &   90 &  61           &  39           \\
\hline
ELG  &  808 &  87           &  46           \\
non--ELG  & 2990 &  39           &  52           \\
\hline
\hline
\end{tabular}
\caption{\label{e1e2tab} 
Distribution with respect to the first and second Principal
Components, ${e_{1}}$ and ${e_{2}}$.  In the upper half of the Table
the galaxies are grouped according to morphology (from D80).  In the
lower half, galaxies are grouped according to the presence (ELG) or
absence (non--ELG) of emission lines.}
\end{table}

\section{Results}
\label{results}

\subsection{Segregations in PCA--space}

The PCA is completely `self-propelled', i.e., it does not need to be
tuned or trained.  Therefore it is interesting to look at the results
of the PCA for the galaxies with morphology from D80, as well as for
the galaxies with and without emission lines, to see in what way the
PCs and the morphology or emission--line character correlate.

In Fig.~\ref{e1e2_1} we show, for all 270 galaxies with morphology
from D80, the distribution with respect to the first and second PCs
for the 3 classes E, S0 and S+I.  All galaxies have values of
${e_{1}}$ between --35 and 35 while ${e_{2}}$ is almost always in the
range [--10,10].  Yet, galaxies of different morphological type have
(slightly) different distributions in the (${e_{1}},{e_{2}}$)-plane.
The ellipticals have predominantly negative values of ${e_{1}}$, the
lenticulars are more evenly spread in ${e_{1}}$ while the spirals and
irregulars have more positive values of ${e_{1}}$ than negative ones.
These differences between the distributions are quantified in Table
\ref{e1e2tab}, which gives the fraction of galaxies with positive
values of ${e_{1}}$ and ${e_{2}}$ for the three morphological classes.
It is clear that the fraction of galaxies with ${e_{1}>0}$ increases
towards later type. There is a tendency for ${e_{2}}$ to be slightly
negative on average, although the effect is not very significant in
view of the statistics. As the information content of the PCs e$_i$
decreases with increasing i, the higher-order PCs will not have, by
themselves, more discriminating power than ${e_{2}}$.

The effect visible in Fig.~\ref{e1e2_1} and Table \ref{e1e2tab} is
qualitatively similar to that found by Lahav et al. (1996) who used 13
galaxy parameters (e.g. blue minus red colour, central surface
brightness) and found that different morphological types occupy
distinct regions in the ${(e_{1},e_{2})}$--plane.  They even detected
a slight separation between E and S0 galaxies, although the regions
occupied by the two morphological types had considerable overlap.

\begin{figure}[!b]
\psfig{figure=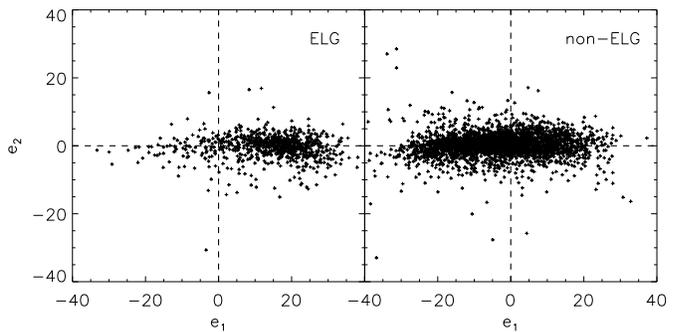,width=8.0cm}
\caption{\label{e1e2_2}
Distribution with respect to the first and second Principal
Components, ${e_{1}}$ and ${e_{2}}$, for emission-line galaxies (ELG,
808 galaxies) and galaxies that do not show emission lines (non--ELG,
2990 galaxies).}
\end{figure}

In Fig.~\ref{e1e2_2} we show the distribution with respect to the
first and second PC for the ELG (lefthand panel) and the non-ELG
(righthand panel). There is a clear difference between the two
distributions, with almost 90\% of the ELG having a positive value of
${e_{1}}$, while the non-ELG have, on average, a slightly negative
value of ${e_{1}}$ (see also Table \ref{e1e2tab}).  Qualitatively,
Figs.~\ref{e1e2_1} and \ref{e1e2_2} are quite consistent, in view of
the fact that almost all ELG are spirals (see Paper III).  It is
interesting to note that apparently the ELG represent those spirals
that have essentially only positive values of ${e_{1}}$.

Note also that the difference between ELG and non--ELG persists if we
do not include in the PCA the spectral ranges where the main emission
lines can occur. This shows that it is not only the emission lines
themselves which distinguish ELG from non--ELG, but that more global
properties of the spectrum, such as continuum slope (see
Sect.~\ref{meaning}), correlate with the presence of emission lines.

\subsection{Success rates}
\label{sucrat}

\begin{table*}
\centering
\caption{\label{tabresults}
Success rates (in percentages) for classifying galaxies with the ANN.
The numbers are averages ($\pm$ {\em r.m.s.} values around these
averages) over 10 realizations of the ANN.  For each realization,
different sets of galaxies (which are partly correlated) are used to
train the ANN.  The first column gives the galaxy type as classified
by D80.  The second column gives the number of galaxies of this type
that is used in the training set.  The third, fourth and fifth columns
give the fraction of galaxies (per morphological type) that is labeled
as E, S0 and S+I, respectively, in the training set by the ANN.
Column 6 gives the number of galaxies in the test set.  Columns 7 to 9
give the fraction of galaxies (per morphological type) that is labeled
as E, S0 and S+I, respectively, in the test set by the ANN.  In the
lower half of the table, the E and S0 galaxies are combined.}

\begin{tabular}{lccccccccc}
\hline
\multicolumn{10}{c}{Three--class system} \\
\hline
type & \multicolumn{4}{c}{Training set} & & \multicolumn{4}{c}{Test set}\\
\cline{2-5} \cline{7-10}
 & N & \%E & \%S0 & \%S+I & & N & \%E & \%S0 & \%S+I \\
\hline
E   & ${53\pm3}$ & ${53\pm19}$ & ${45\pm19}$ & 
 ${ 2\pm2}$ & & ${ 9\pm3}$ & ${34\pm28}$ & 
 ${50\pm24}$ & ${16\pm12}$ \\
S0  & ${55\pm3}$ & ${13\pm 8}$ & ${68\pm10}$ & 
 ${19\pm6}$ & & ${63\pm3}$ & ${23\pm11}$ & 
 ${49\pm15}$ & ${28\pm 7}$ \\
S+I & ${55\pm3}$ & ${ 1\pm 1}$ & ${18\pm 6}$ & 
 ${81\pm5}$ & & ${34\pm3}$ & ${13\pm 7}$ & 
 ${24\pm 8}$ & ${63\pm 6}$ \\
\hline
\hline
     &     &     &      &       & &     &     &      &       \\
\hline
\multicolumn{10}{c}{Two--class system} \\
\hline
type & \multicolumn{4}{c}{Training set} & & \multicolumn{4}{c}{Test set}\\
\cline{2-5} \cline{7-10}
 & N & \multicolumn{2}{c}{\%E+S0} & \%S+I & & N & \multicolumn{2}{c}{\%E+S0} & \%S+I \\
\hline
E/S0 & ${108\pm 5}$ & \multicolumn{2}{c}{${90\pm3}$} & 
 ${10\pm3}$ & & ${72\pm5}$ & 
 \multicolumn{2}{c}{${78\pm6}$} & ${22\pm6}$ \\
S+I  & ${ 55 \pm3}$ & \multicolumn{2}{c}{${19\pm5}$} & 
 ${81\pm5}$ & & ${34\pm3}$ & 
 \multicolumn{2}{c}{${37\pm6}$} & ${63\pm6}$ \\
\hline
\hline
\end{tabular}
\end{table*}

In Table \ref{tabresults} we give the results of our morphological
classification using the ANN operating on the 15 most significant PCs.
The percentages quoted are averages ($\pm$ {\em r.m.s.} values around
these averages) over 10 realizations of the ANN. For each realization,
different sets of galaxies (which are thus partly correlated) are used
to train the ANN. We give this information for the {\em test set} as
well as for the {\em training set}. We consider two cases: the
three--class ANN classification (top), and its compressed pseudo
two--class version obtained by combining the E and S0 classes of the
three--class classification (see Sect.~\ref{training}).

The overall success rates for the training and test sets are
${67\pm5\%}$ and ${49\pm6\%}$, respectively, for the three--class
system, giving each of the three classes equal weight. The success
rate for the training set is larger than for the test set, which must
be due to the fact that the ANN weights are calculated using the
galaxies in the training set only. The success rate for the test set,
however, is the one that should be applied to the entire set of
galaxies to be classified.

If one uses the two--class system, viz. separating between early-- and
late--types only, the success rate for the training set is
${87\pm3\%}$ and for the test set ${73\pm4\%}$.  Obviously, these
success rates are larger than for the three--class system because one
has less categories to classify the galaxies in, and because a large
fraction of the classification `failures' occurs between E's and
S0's. The fact that in the two--class system the success rate of the
early--type galaxies is higher than for the late--type galaxies may be
due, at least partly, to the asymmetry between early-- and late--type
galaxies in our spectral misclassifications, as discussed in
Sect.~\ref{misclass}.

Using the galaxies classified by Dressler, we determined how the
spirals that are incorrectly classified are distributed between
early-- and late--type spirals. For the training set, only 30\% of the
spiral galaxies that are classified as S0 by our ANN are of type Sb or
later, according to D80.  Sodr\'e \& Cuevas (1997) obtained a similar
result, namely that the spectral variation, as measured by the first
PC, is slow from E to Sab and increases strongly for later types.  So
one expects few Sb or later--type spirals to be classified as
early--type.  For the test set, all spirals classified as E by the ANN
are of type Sa.  For the training set 30\% of the spirals of type Sa,
10\% of type Sb, 13\% of type Sc and 0\% of type Sd+I are classified
incorrectly.  For the test set, these numbers are 26\% for Sa, 13\%
for Sb and 0\% for Sc or later.  The early--type spirals are thus more
often misclassified as E or S0 than the late--type spirals.

The fraction of the 808 ELG in the ENACS sample, used in the present
analysis, that is classified as E, S0 or S+I is ${7\pm5\%}$,
${13\pm4\%}$ and ${80\pm7\%}$, respectively.  Biviano et al.  found
that out of the 71 ELG with a morphological type available, 86\% are
of type S+I, 11\% are S0 and 3\% are elliptical.  The fraction of ELG
that is classified as spiral (80\%) is higher than would be expected
from the individual success rates for the E, S0 and S+I subsamples
(Table \ref{tabresults}) and the distribution of ELG over
morphological type (Paper III).  These would imply that
$[0.03\times(16\pm12)] + [0.11\times(28\pm7)] +
[0.86\times(63\pm6)] =$ $58\pm5\%$ of the ELG would be classified as a
spiral.  Apparently, the success rate for the ELG is larger than for
the entire data set containing both ELG and non--ELG, which could
imply that ELG are preferentially {\em late} spirals, for which the
classification is more reliable than it is for early spirals.

Based on the results of Table \ref{tabresults} one expects ${59\pm32}$
E's, ${110\pm35}$ S0's and ${100\pm9}$ S+I galaxies in the set of 270
ENACS spectra with a classification by D80.  These numbers agree very
well with the actual numbers in the D80 set, viz. 62 E's, 118 S0's and
90 S+I.  However, this is not too surprising, as the correspondence
between both sets of numbers was one of our criteria to set the output
ranges (Sect.~\ref{optres}).

The distribution of galaxy types for the entire sample of 3798 spectra
in our final sample is: ${24\pm5\%}$ E, ${33\pm4\%}$ S0
and ${43\pm7\%}$ S+I.

Of all AGN in our sample, ${55\pm10\%}$ is classified as early--type.
This is significantly more than the ${20\pm7\%}$ of {\em all} ELG that
is classified as early--type.  Apparently, there are significantly
more early--type galaxies among AGN than there are among the non--AGN
ELG.

In Table \ref{tabresults2} we give the success rates for the galaxy
classification if one uses different numbers of PCs. It appears that
the results obtained with 10 PCs in the ANN may be marginally worse
than those with 15 or 20 PCs, but the differences are not very
significant.

\begin{table}
\begin{tabular}{lccc}
\hline
\multicolumn{4}{c}{Training set} \\
\hline
system       & 10 PCs & 15 PCs & 20 PCs \\
\hline
three--class & ${62\pm4}$ & ${67\pm5}$ & ${69\pm4}$ \\
two--class   & ${85\pm2}$ & ${87\pm3}$ & ${86\pm3}$ \\
\hline
\hline
             &        &        &        \\
\hline
\multicolumn{4}{c}{Test set} \\
\hline
system       & 10 PCs  & 15 PCs & 20 PCs \\
\hline
three--class & ${43\pm 7}$ & ${49\pm6}$ & ${48\pm6}$ \\
two--class   & ${68\pm13}$ & ${73\pm4}$ & ${70\pm6}$ \\
\hline
\hline
\end{tabular}
\caption{\label{tabresults2}
Success rates (in percentages) for classifying galaxies with the ANN
using different numbers of Principal Components (PCs).  The numbers
are averages ($\pm$ {\em r.m.s.} values around these averages) over 10
realizations of the ANN.  For each realization, different sets of
galaxies (which are partly correlated) are used to train the ANN.  The
results are given for the training and test sets separately and
both for the three--class and the two--class classification systems.}
\end{table}

We have also run the PCA and ANN with the spectra of the September
1992 period included as well.  The classification results then are
${63\pm3\%}$ (three--class) and ${83\pm2\%}$ (two--class) for the
training set, and ${44\pm3\%}$ (three--class) and ${69\pm2\%}$
(two--class) for the test set.  These success rates are slightly
lower than those if the spectra from the September 1992 period are not
included, and they justify our choice not to include those galaxies in
the analysis.

Finally, we have investigated if the success rates depend clearly on
the S/N--ratio of the galaxy spectrum. This is not the case, as is
expected because, by construction, the first PCs will contain
relatively little noise.

\section{Spatial and Kinematical Differences between Early-- and 
Late--type Galaxies}
\label{spatkin}

Biviano et al. (1997) studied the differences between ELG and non--ELG
as far as their spatial distribution and kinematical properties are
concerned.  Combining the data for 75 clusters with at least 20 member
galaxies, they found that the line--of--sight velocity dispersion
(with respect to the cluster mean velocity), ${\sigma_{los}}$, is ${21
\pm 2\%}$ larger for the ELG than it is for the non--ELG. They also 
found that the spatial distribution of the ELG is significantly less
peaked towards the cluster centre than that of the non--ELG. 

For a full appreciation of this result it is important to remember
that the subsample of ELG consists almost exclusively of late--type
galaxies, whereas the subset of non--ELG contains galaxies of all
types. In other words: if the late--type galaxies without
emission-lines would share the distribution and kinematics of their
ELG counterparts, the differences between early-- and late--type
galaxies could well be more pronounced than between non-ELG and ELG.

On the other hand, it is also quite possible that the less
centrally--concentrated distribution and larger ${\sigma_{los}}$ apply
only to the late-type galaxies {\em with} emission lines. If so, that
would provide additional support for the conclusion in Paper III that
the ELG are likely to be on fairly radial, first--approach orbits, as
suggested by their larger velocity dispersion, their projected spatial
distribution, and their rather steep velocity dispersion profile
${\sigma_{los}(r_{proj})}$.  The presence of the line--emitting gas
would be fully consistent with this picture.

\subsection{Kinematics}

We have repeated part of the analysis of Paper III, making use of the
classification in early-- and late--type galaxies on the basis of the
spectrum, discussed in this Paper. We start with the same set of 75
clusters as in Paper III.  However, our galaxy sample includes only
those galaxies for which we could estimate the galaxy morphology from
the PCA and ANN.  This limits the sample to 2594 galaxies in 66
clusters, of which 399 galaxies are ELG, while ${1571\pm52}$ are
classified as early--type, and ${1023\pm52}$ as late--type.  For each
galaxy the normalized line--of--sight component of the velocity
w.r.t. the cluster centre, ${v_{norm} = (v-v_{clus})/\sigma_{clus}}$,
is determined , where ${v_{clus}}$ is the average cluster velocity and
${\sigma_{clus}}$ is the cluster line--of--sight velocity dispersion
of the cluster to which the galaxy belongs. Following Paper III, we
construct one large composite cluster by combining the data of all 66
clusters.

Using this sample of 2594 galaxies in 66 clusters, we find that the
normalized line--of--sight velocity dispersion ${\sigma_{los}}$ of the
ELG is 23\% larger than that of the non-ELG, which is fully consistent
with the result of Paper III. The values of ${\sigma_{los}}$ for ELG
and non--ELG are given in column 3 of Table \ref{kin_tab}. The value
of ${\sigma_{los}}$ for the dominant class of non--ELG is larger than
unity because, in constructing the composite cluster, one adds
velocity distributions for which the average velocities are known only
with a limited accuracy. This leads to the superposition of
(approximately Gaussian) velocity distributions with small apparent
offsets, which slightly increases the dispersion above the expected
value of 1.00. As discussed extensively in Paper III, this effect
certainly does {\em not} explain the value of ${\sigma_{los}}$ of 1.28
for the ELG, because there is no evidence that the ELG have
significant velocity offsets w.r.t. the non-ELG.

\begin{table}
\centering
\caption{\label{kin_tab}
Line--of--sight velocity dispersion (with respect to the cluster
centre) and parameter values of the best--fitting
${\beta}$--model ${\Sigma(r) = \Sigma(0) \left[
1+(r/r_{c})^{2} \right] ^{\beta}}$ to the surface density profiles of
galaxies.  Column 1 gives the subsample of galaxies.  Column 2 gives
the number of galaxies in this subsample.  All values are averages
($\pm$ {\em r.m.s.} values around these averages) over 10 realizations
of the ANN.  The best--fitting model parameters are not listed for the
early--type ELG, as these are very uncertain.}
\begin{tabular}{lcccc}
\hline
sample         & N                    & ${\sigma_{los}}$ & ${\beta}$ & ${r_{c}}$ \\
\hline
all            & 2594                 & 1.08 & --0.66 & 0.10 \\
               &                      &      &        &      \\
non--ELG       & 2195                 & 1.04 & --0.69 & 0.10 \\
ELG            &  399                 & 1.28 & --0.67 & 0.27 \\
               &                      &      &        &      \\
early--type    & ${1571\pm52}$ & ${1.03\pm0.01}$ & --0.66 &0.08\\
late--type     & ${1023\pm52}$ & ${1.15\pm0.02}$ & --0.76 &0.23\\
               &                      &      &        &      \\
early--type    &                      &      &        &      \\
\hspace*{1.1cm} non--ELG  & ${1504\pm51}$ & ${1.02\pm0.01}$ & --0.70 &0.09\\
\hspace*{1.1cm} ELG       & ${67\pm24}$   & ${1.22\pm0.14}$ & --     & -- \\
               &                      &      &        &      \\
late--type     &                      &      &        &      \\
\hspace*{1.1cm} non--ELG  & ${691\pm51}$  & ${1.09\pm0.03}$ & --0.76 &0.20\\
\hspace*{1.1cm} ELG       & ${332\pm24}$  & ${1.28\pm0.01}$ & --1.24 &0.71\\
\hline
\hline
\end{tabular}
\end{table}

In Table \ref{kin_tab} we also give the values of ${\sigma_{los}}$ for
several other subsets of the total sample. It appears that the
${\sigma_{los}}$ of the late--type galaxies is ${12\pm3\%}$ larger
than that of the early--type galaxies. This difference is
significantly smaller than it is for ELG versus non--ELG, which makes
it unlikely that the non-ELG spirals have the same kinematics as the
ELG (mostly late spirals). This is indeed confirmed by the value of
${\sigma_{los}}$ for the non-ELG late-type galaxies (mostly early
spirals) of $1.09\pm0.03$. Although this is somewhat higher than the
value of 1.04 for all non-ELG, it is also very much smaller than the
value of 1.28 found for all ELG, and for the subset of late-type ELG.

The intermediate value of $1.09\pm0.03$ for the non-ELG late-type
galaxies may mean one of three things. First, and most simply, it may
be a statistical fluke, i.e. a 2$\sigma$ excursion of a value that is
not fundamentally different from the $1.03\pm0.01$ that we find for
the early-type galaxies. Secondly, the separation of the late-type
galaxies into ELG and non--ELG may not be perfect. This could be a
result of our observational limit for the detection of emission lines,
which need not correspond exactly to a kinematical distinction. In
other words: the non--ELG late--type galaxy category may contain a
fraction (which must be significant) of intrinsic ELG, for which the
emission lines were not detectable in the ENACS. In that case, the
true ${\sigma_{los}}$ of the non-ELG late--type galaxies is smaller
and closer to the value of $1.03\pm0.01$ found for the early--type
galaxies. Thirdly, the non-ELG late--type galaxies may be a
dynamically `pure' class, with kinematics intermediate between that of
the early-type galaxies and that of the late--type ELG.

One might have a slight worry that the results in Table \ref{kin_tab}
are somewhat influenced by the fact the separation between e.g. early-
and late-type galaxies on the basis of the spectrum is not perfect. In
other words: the value of ${\sigma_{los}}$ for the early-type class
may have been somewhat overestimated because the early-type class
contains a non-negligible contribution of late-type galaxies.
Similarly, the value of ${\sigma_{los}}$ for the late-type class may
be somewhat underestimated. However, these effects are small.

Using the success rates in Table~\ref{tabresults} for the two-class
system, we estimate that at most 1 out of 4 galaxies in the early-type
class is a misclassified late-type galaxy. Because essentially all
galaxies in the early-type class (i.e. including the misclassified
late-type galaxies) are non-ELG, the value of ${\sigma_{los}}$ of the
early-type class is not overestimated very much. Using the value of
${\sigma_{los}}$ of 1.09 for the late-type non-ELG galaxies (which is
a slight underestimate, see below), we estimate the bias in
${\sigma_{los}}$ of the non-ELG early-type galaxies to be at most a
few percent. With this result, we can estimate that the value of
${\sigma_{los}}$ of the late-type non-ELG is more likely to be about
1.13 rather than 1.09, but this is still considerably smaller than the
value of 1.28 of the late-type ELG.

Therefore, the data in Table \ref{kin_tab} support a picture in which
there is a clear correlation between the presence of emission lines
and a high velocity dispersion. Rather unexpectedly perhaps, the ratio
between the ${\sigma_{los}}$ of ELG and that of non--ELG does not
appear to depend on whether the ELG or non--ELG are early-- or
late--type galaxies. The ELG among the early-- and late--type galaxies
have a value of ${\sigma_{los}}$ that is about 18\% larger than that
of the non--ELG of the corresponding galaxy type. In view of the large
uncertainty in the estimate of ${\sigma_{los}}$ for the early--type
ELG, this may be totally fortuitous, however, and we certainly should
not overinterpret this result.

In summary, the basic factor driving the difference in kinematics
seems to be the presence or absence of emission lines, whereas the
distinction between early-- and late--type galaxies is less important,
while the class of late--type non--ELG presents an intrigueing
cross-breed which may hold important clues to the physical meaning of
the results.

\subsection{Projected distributions}

\begin{figure*}
\psfig{figure=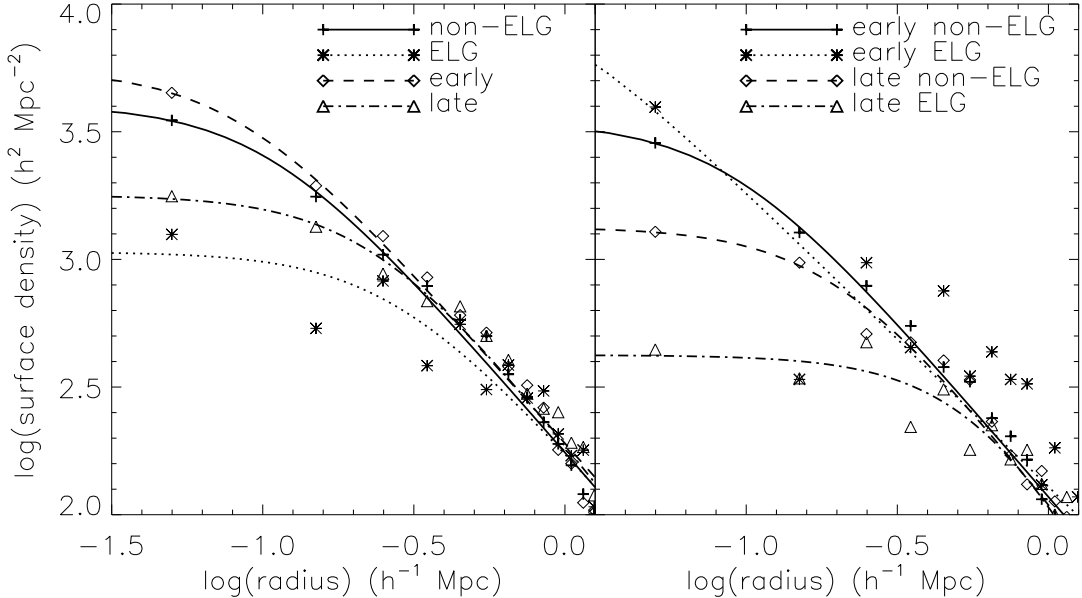,width=14.0cm}
\caption{\label{dens_prof}
Surface density profiles for the various subsamples of galaxies.  The
lines are the best--fitting ${\beta}$--models.  The parameters
of these are listed in Table 5.}
\end{figure*}

In view of the results in Table \ref{kin_tab} and in Paper III, it is
interesting to see how the kinematics and the projected spatial
distribution are related. We therefore determined the surface density
profiles of all subsamples, which we show in Fig.~\ref{dens_prof}.
The profiles are averages over the 10 realizations of the ANN (for the
samples based on the distinction between early-- and late--type). The
profiles are shifted vertically such that at ${r=1 h^{-1}}$ Mpc the
fitted profiles have the same values.  The lines show the
best--fitting ${\beta}$--model,
\begin{eqnarray}
{\Sigma(r)} & {=} & {\Sigma(0) \left[ 1+(r/r_{c})^{2} 
\right] ^{\beta}}
\end{eqnarray}
where $\Sigma$ is the surface density, r$_c$ is the core--radius and
$\beta$ is the logarithmic slope at large radii. The best--fitting
values of ${\beta}$ and ${r_{c}}$ are given in columns 4 and 5 of
Table \ref{kin_tab}. Note that, as a result of the details of the
OPTOPUS observations, the spatial completeness of the galaxy samples
may not be uniform, so that the estimate of ${\beta}$ may be
biased. The errors in the estimates, determined from the comparison of
the 10 realizations of the ANN, are small, of the order of 10\%.  Only
for the early--type ELG the errors are substantially larger because
the number of galaxies in this subsample is small.

The non--ELG are significantly more centrally peaked than the ELG, as
was already concluded by Biviano et al.  Although we find the same
value of ${\beta}$ for both subsamples, the ELG population has a much
larger core--radius ${r_{c}}$ than that of the non-ELG.  The
difference between early-- and late--types is similar to that between
non--ELG and ELG, the former being more centrally concentrated than
the latter.  The subsample of late--type ELG has a value of ${\beta}$
that seems different from that of all other subsamples, but the
difference probably is not significant, as ${r_{c}}$ is quite
large. 

Apparently, the late--type ELG are distributed much more towards the
cluster outskirts than all other galaxies, including the late--type
non--ELG.  For the early--type ELG, the values of ${\beta}$ (--0.58)
and ${r_{c}}$ (0.02) are not very reliable because of the small number
of galaxies involved.  However, from a comparison between all ELG and
the late--type ELG, one may conclude that both ${\beta}$ and ${r_{c}}$
are probably quite small for the early--type ELG.  So the distribution
of early--type ELG probably also deviates from that of the other
galaxies, in the sense that they are more centrally concentrated.  As
we have seen in Sect.~\ref{sucrat}, the early--type ELG are often AGN,
and this result therefore is not too surprising. 

However, the early-type ELG may also contain a contribution from
central dominant galaxies with emission lines from cooling flows
(e.g. Heckman et al. 1989, and Crawford et al. 1995), which might give
an important contribution to the high surface density of early-type
ELG in the innermost bin in Fig.~\ref{dens_prof}. Yet, it is not clear
that the line ratios of the lines we observe are consistent with this
explanation, and from our present data it is not easy to estimate
this contribution.

\subsection{What does it mean?}

Combining the results of the spatial and kinematical properties of the
different galaxy populations, we conclude that the late--type non--ELG
have properties that resemble more those of the early--type galaxies,
i.e. most of the other non--ELG. Yet, their projected distribution is
slightly wider than that of the early--type galaxies, with a core
radius that is a factor two to three larger, and kinematically they
are somewhat `hotter' than the other non-ELG. The (late--type) ELG,
which consist mostly of spirals, behave very differently.  Their
line--of--sight velocity dispersion ${\sigma_{los}}$ is much larger
than that of the late--type non--ELG and they are located more towards
the outer regions of clusters.

In Paper III the kinematical characteristics of ELG and non-ELG were
interpreted as an indication for the ELG to be mostly on fairly (but
not necessarily purely) radial orbits, in contrast to the
non-ELG. Combining this with the larger velocity dispersion of the ELG
and their relative scarcity in the very central regions of the
clusters, we were led to the hypothesis that the ELG are mostly on
radial, first-approach infall orbits towards the central regions of
their clusters. This would be consistent with the presence of the
line-emitting gas, as it is likely that that would have been removed
from the galaxy on traversing the dense cluster core.

When the ELG are on orbits which are sufficiently radial without,
however, traversing the very central regions, it is possible that they
have already made several crossings without losing their gas, and will
continue to do so until they `get caught'. In other words: their high
velocity dispersion need not necessarily imply `first approach'
orbits, because in the absence of an encounter they could maintain
their velocity, which was due to their `late' infall. We may assume
that an ELG which gets too close to the cluster centre (either on its
first approach or after several crossings) will probably be
`converted' almost instantly into a non--ELG late--type galaxy, as the
gas gets stripped. 

How the gas gets stripped from the ELG is not totally clear. In
principle, ram pressure against intracluster gas could do the
trick. However, that probably would not change the kinematics and
distribution of the left-overs as drastically as observed.
Alternatively, the harassment of galaxies through fairly high-speed
and relatively distant encounters, as described by Moore et
al. (1996), could be responsible for driving out the gas. Such
encounters could be sufficiently frequent (about once per Gyr) to
ensure that gas-rich ELG are virtually absent from the central
regions. Actually, it is possible that an ELG has to experience a few
of those encounters to get rid of its gas.

However, it is not immediately clear that such encounters will
`instantly' reduce their velocity dispersion of 1.28 to 1.09, the
value observed for their non-ELG counterpart. One factor which may
contribute to this large apparent reduction is projection.  If the ELG
are indeed or fairly radial orbits, and their gas-robbed encounter
products are on less radial orbits, this geometric effect might be
responsible for most of the apparent reduction of ${\sigma_{los}}$.

Thus, it is possible that a slight change of the orbit characteristics
(in particular the anisotropy parameter $\cal A$), which results from
the encounter which strips the ELG from its gas, is sufficient to
considerably reduce ${\sigma_{los}}$ and produce a more centrally
concentrated distribution. Note, however, that the kinematics and
distribution of these stripped ELG may be different from those of the
early-type galaxies which suggests that the latter are a more advanced
product of encounters in the central cluster region.

The AGN among the ELG are a special class. They are predominantly
ellipticals which, probably because of their central location, show
AGN characteristics. Our data unfortunately do not allow us to
determine convincingly how their velocity dispersion compares to those
of the other types of galaxies in the cluster (see Table
\ref{kin_tab}), but they seem to be at least as centrally concentrated
as the non--AGN early--type galaxies.

\section{Summary and Conclusions}
\label{conclusions}

We studied the spatial and kinematical properties of early-- and
late--type galaxies in a subset of the rich Abell clusters observed in
the ENACS.  We compared these properties for galaxies with emission
lines (ELG) and without (non--ELG).

As for only about 10\% of the galaxies in the ENACS the morphological
type was known from imaging, we applied a Principal Component Analysis
(PCA) in combination with an Artificial Neural Network (ANN), to try
and classify all galaxies in the ENACS on the basis of their spectrum.
The PCA is an important first step in the classification as it
compresses essentially all significant information of a spectrum into
a limited number of Principal Components (PCs).  

These PCs, which are linear combinations of the original spectral
fluxes, were subsequently used in an ANN.  The ANN was trained to
classify spectra using a subset of 150 galaxies for which the
morphological type is known (D80).  Another 120 galaxies for which the
morphology is available from D80, the so--called test set, were used
to determine the success rate of the classification algorithm.

Classifying galaxies into three classes (E, S0 and S+I), the ANN
yielded the correct galaxy morphology of D80 for ${49\pm6}$\% of the
galaxies in the test set.  The success rate increased to
${73\pm4}$\% when the galaxies were classified into only two classes,
early-- (E+S0) and late--type (S+I) galaxies.  Furthermore,
${80\pm7}$\% of the galaxies with emission lines in their spectrum
(ELG) was classified as late--type.  This fraction is larger than the
${58\pm5}$\% that one expects from the individual success rates for
each morphological type separately. Apparently, the success rate for
the ELG is larger than for the entire set of galaxies.

We discussed several factors that may produce misclassifications.
First, one does not always know the {\em true} galaxy type. Using
galaxies which Dressler classified twice, we estimate that at least
between 5 and 10\% of the classifications based on imaging are
incorrect.  Secondly, even within one morphological type the spectra
may be substantially different. Thirdly, S0 galaxies may be hard to
separate from ellipticals from their spectrum alone, and we find that
a rather large number of E's is classified as S0 and vice versa,
whereas the number of misclassifications between S0 and S+I is much
smaller. Finally, spiral galaxies with a large bulge may have a
spectrum that leads to a bona-fide early--type classification with PCA
and ANN.

We investigated how galaxies of different type are distributed in the
plane defined by the two most significant PCA components. There
appears to be a distinction between E, S0 and S+I galaxies in this
plane, although it is not very large. On the contrary, the ELG and
non--ELG have clearly different distributions, which shows that the
PCs contain significant information about the morphological type of a
galaxy.

Finally, we extended the analysis of Biviano et al. (in Paper III),
who studied the differences in the spatial and kinematical properties
of ELG and non--ELG, to galaxies of different morphology.  We conclude
that the presence of emission lines, rather than the galaxy
morphology, is the basic property that is correlated with the spatial
and kinematical properties of a galaxy.  Thus, the correlation between
morphology on the one hand and spatial and kinematical properties on
the other hand seems to result mainly, if not exclusively, from the
presence of emission lines.

The line--of--sight velocity dispersion with respect to the average
cluster velocity is larger for the ELG than it is for the non--ELG,
and. A similar, but smaller, difference is found between late-- and
early--type galaxies.  This supports the idea that the ELG are on
fairly radial, and possibly `first approach' orbits towards their
cluster centres, while their line--emitting gas has not yet been
stripped. In addition, the late--type galaxies {\em without} emission
lines (i.e. with little gas) appear to have spatial and kinematical
properties that more resemble those of the early--type galaxies than
those of the late--type galaxies {\em with} emission lines.
Apparently, if a late--type galaxy has passed through the cluster
centre, most of its gas will have been stripped and the galaxy will
not show emission lines anymore.

\begin{acknowledgements}

We thank Simon Folkes, Ofer Lahav and Avi Naim for valuable
discussions and suggestions.  Ben Kr\"ose is acknowledged for a
stimulating discussion, and Richard Arnold for triggering our interest
in Artificial Neural Networks. We thank the ENACS team for allowing us
to use the data, and Jaime Perea in particular for comments about the
spectral reduction.

\end{acknowledgements}

\vfill
\end{document}